\documentclass[journal,twoside,web]{ieeecolor}
\usepackage{generic}
\usepackage{cite}
\usepackage{amsmath,amssymb,amsfonts}
\usepackage{algorithmic}
\usepackage{graphicx}
\usepackage{algorithm,algorithmic}
\usepackage{hyperref}
\hypersetup{hidelinks=true}
\usepackage{textcomp}
\usepackage{booktabs} 
\usepackage{tabularx} 
\usepackage{caption}  
\usepackage{verbatim}
\usepackage{graphicx}
\usepackage{amsmath}
\usepackage{lineno,hyperref}
\usepackage{multirow}
\usepackage{stfloats}
\usepackage{newtxmath}   
\usepackage{arydshln}
\usepackage{array} 
\usepackage{makecell}   
\usepackage{threeparttable}

\def\BibTeX{{\rm B\kern-.05em{\sc i\kern-.025em b}\kern-.08em
    T\kern-.1667em\lower.7ex\hbox{E}\kern-.125emX}}
\begin{document}
\title{HCFT: Hierarchical Convolutional Fusion Transformer for EEG Decoding }
\author{Haodong Zhang, Jiapeng Zhu, Yitong Chen, Hongqi Li*, \IEEEmembership{Member, IEEE}
\thanks{Manuscript received Nov 12, 2025. This work was supported in part by the Natural Science Basic Research Program of Shaanxi Province under Grant 2024JC-YBQN-0659 (Corresponding author: Hongqi Li.)  }
\thanks{H. Zhang, J. Zhu, Y. Chen, and H. Li are with the School of Software, Northwestern Polytechnical University, Xi’an 710072, China (e-mail: zhang\_haodong@mail.nwpu.edu.cn, zhujp@mail.nwpu.edu.cn, chenyt@mail.nwpu.edu.cn, lihongqi@nwpu.edu.cn).}}

\maketitle
\begin{abstract}
Electroencephalography (EEG) decoding requires models that can effectively extract and integrate complex temporal, spectral, and spatial features from multichannel signals. To address this challenge, we propose a lightweight and generalizable decoding framework named Hierarchical Convolutional Fusion Transformer (HCFT), which combines dual-branch convolutional encoders and hierarchical Transformer blocks for multi-scale EEG representation learning. Specifically, the model first captures local temporal and spatiotemporal dynamics through time-domain and time-space convolutional branches, and then aligns these features via a cross-attention mechanism that enables interaction between branches at each stage. Subsequently, a hierarchical Transformer fusion structure is employed to encode global dependencies across all feature stages, while a customized Dynamic Tanh normalization module is introduced to replace traditional Layer Normalization in order to enhance training stability and reduce redundancy. Extensive experiments are conducted on two representative benchmark datasets, BCI Competition IV-2b and CHB-MIT, covering both event-related cross-subject classification and continuous seizure prediction tasks. Results show that HCFT achieves 80.83\% average accuracy and a Cohen’s kappa of 0.6165 on BCI IV-2b, as well as 99.10\% sensitivity, 0.0236 false positives per hour, and 98.82\% specificity on CHB-MIT, consistently outperforming over ten state-of-the-art baseline methods. Ablation studies confirm that each core component of the proposed framework contributes significantly to the overall decoding performance, demonstrating HCFT’s effectiveness in capturing EEG dynamics and its potential for real-world BCI applications.
\end{abstract}


\section{Introduction}
\label{sec:I}
\IEEEPARstart{E}{LECTROENCEPHALOGRAM} decoding is a critical component of modern brain-computer interfaces (BCIs), enabling applications in medical rehabilitation, affective computing, and human-machine interaction \cite{1,2,3,4}. While deep learning has markedly advanced the field, models based on Convolutional Neural Networks (CNNs) and Recurrent Neural Networks (RNNs) are inherently constrained by their local receptive fields, struggling to capture the long-range, global dependencies that are crucial for accurate neural representation. Instead, the recent advent of the Transformer model, with its powerful self-attention mechanism, offers a promising alternative for modeling these complex spatiotemporal dynamics.

Specifically, Transformer-based architectures offer strong capabilities for modeling long-range dependencies from sequential data, and thus numerous studies have applied them to EEG decoding \cite{5}. Typically, Wang et al. \cite{6} proposed a Transformer-based architecture that hierarchically learns discriminative spatial information to enhance the capture of EEG spatial dependencies, achieving favorable results in emotion recognition tasks. Xie et al. \cite{7} designed five categories of Transformer-based models for different EEG scenarios based on the importance of spatiotemporal dependencies between different channels for accurate classification. The developed models incorporated positional encoding into the Transformer, improving the classification performance and performing well in cross-subject validation. Song et al \cite{8} designed a model called EEG Conformer to decode EEG signals by CNN to extract local features and Transformer to capture the global features, demonstrating excellent performance on the tasks of motor imagery (MI) classification and emotion recognition. Similarly, ACTNN \cite{9} cascades CNN and Transformer, integrating crucial spatial, spectral, and temporal EEG information effectively, achieving outstanding performance on two public emotion recognition datasets. Furthermore, Wei et al. \cite{10} designed a model named TC-Net that employed the Transformer module to capture the global features of EEG, and a new strategy of EEG-PM was used to merge the neighboring patches and better extract local features. It performs well in the subject-dependent scenario of emotion recognition task, while obtaining relatively low performance in the cross-subject recognition task. Siddhad et al. \cite{11} also achieved excellent classification results using Transformer network on raw EEG data without feature extraction, indicating the Transformer-based deep learning network can abate the need for heavy feature-extraction of EEG data. Zheng and Pan \cite{12} designed STS-Transformer for multi-channel EEG without data preprocessing to extract spatio-temporal features automatically, achieving remarkable performance in emotion recognition tasks. A most recently proposed framework of Dual-TSST fully incorporate CNN and Transformer to extract comprehensive spatio-frequency features through a dual-branch architecture, demonstrating remarkable performance in various EEG decoding \cite{13}. 

Despite these promising developments, several key limitations persist. A predominant challenge is the limited generalization capability of these models in cross-subject validation scenarios, indicating a reliance on subject-specific characteristics. Moreover, existing architectures often feature considerable structural complexity, leading to high computational costs. Most critically, there is a frequent lack of effective synergy and explicit modeling of the interrelation-ships between temporal, spatial, and spectral features. These limitations highlight the need for novel decoding frameworks that can simultaneously capture fine-grained temporal rhythms, spatial patterns across electrodes, and multi-scale global dependencies, while maintaining parameter efficiency and ensuring stable training dynamics.

To address the aforementioned limitations, we introduce the Hierarchical Convolutional Fusion Transformer (HCFT), a novel EEG decoding framework. Inspired by the Pyramid Vision Transformer (PVT) \cite{14}, HCFT uses a hierarchical structure that downsamples feature maps to manage computational complexity. The architecture is guided by three core principles: (i) local feature extraction with cross-modal guidance, (ii) modular stacking enhanced for depth stability, and (iii) one-shot multi-scale token fusion.

The model first processes EEG signals through a dual-branch depthwise separable convolutional encoder. One branch extracts one-dimensional temporal features, while the other captures two-dimensional spatiotemporal patterns. A conditional cross-attention mechanism then uses the temporal features as queries to guide the spatiotemporal branch, explicitly aligning fine-grained temporal rhythms with spatial patterns in a computationally efficient manner.

These features are processed by a unified Convolutional Fusion Transformer (CFT) Block, which integrates Multi-Head Self-Attention (MHSA), Multi-Head Cross-Attention (MHCA), and an expanded feed-forward network. To ensure stable training in deep stacks, we optionally incorporate a Dynamic Tanh Normalization (DyT) module \cite{15}, which improves convergence without incurring inference overhead.

Finally, a pyramidal encoder facilitates multi-scale fusion. Tokens from different hierarchical stages are concatenated and integrated via self-attention. A subsequent long-context pooling operation condenses this information into a compact representation, enabling the one-shot integration of high-resolution local features and long-range global dependencies.

The remainder of this paper is organized as follows. Section II presents the architecture and implementation details of the HCFT model. Section III reports dataset and experimental setups. Section IV provides the detailed results with ablation analyses. Finally, section V discusses and concludes the study, with future directions outlined.

\begin{figure*}
    \centering
    \includegraphics[width=\textwidth]{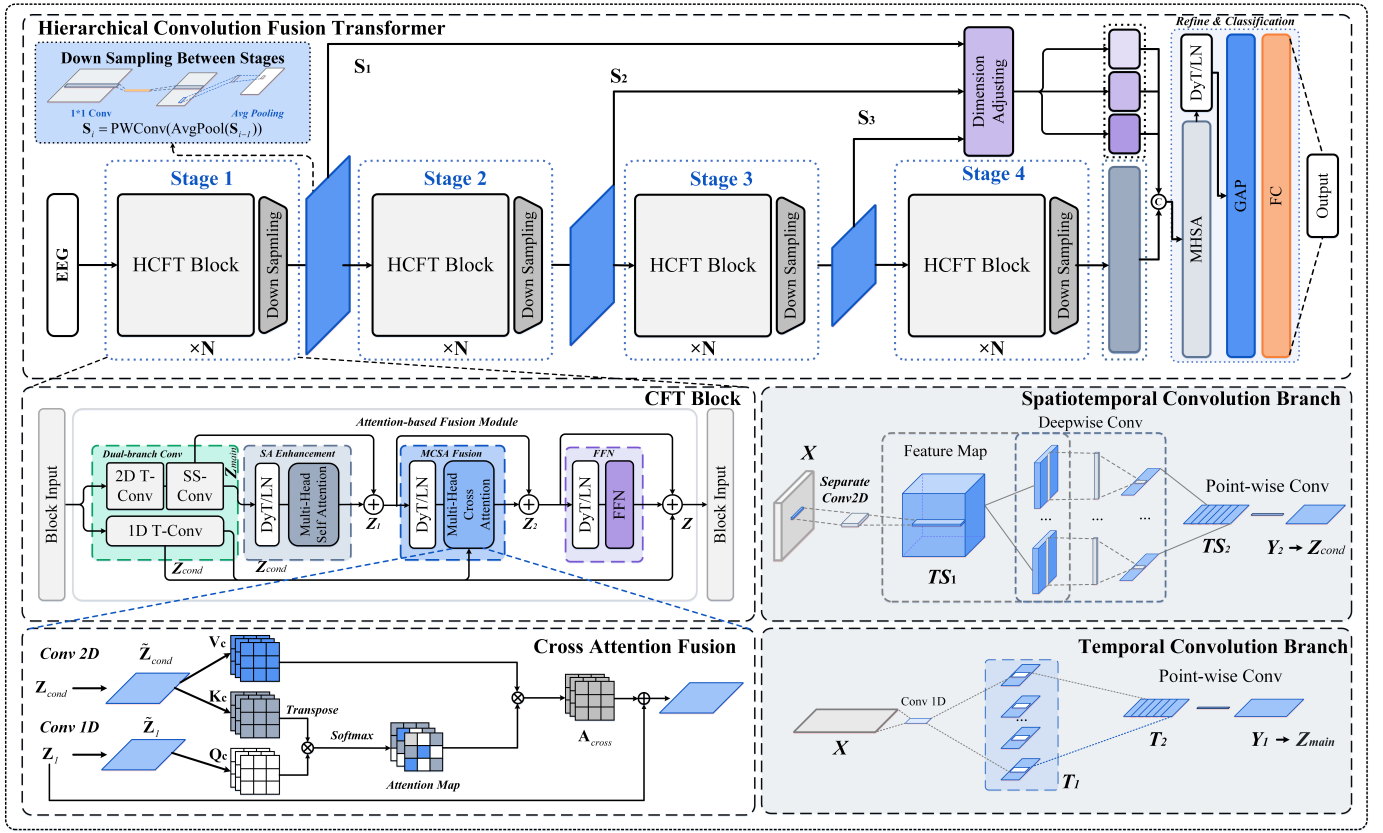}
    \caption{The proposed HCFT framework, include Multiple Stage design with different number of CFT Blocks}
    \label{fig 1}
\end{figure*}

\section{Methodology}
\label{sec:II}
\subsection{Overall Model Architechture}
As demonstrated in Fig.\ref{fig 1}, HCFT adopts a multi-stage hierarchical encoder, which is grouped into progressively deeper semantic stages (i.e., Stage 1, Stage 2, Stage 3, Stage 4) and each stage contains a different number (N1, N2, N3, N4) of cascaded convolution-attention fusion blocks, referred to as CFT Blocks, enabling the whole network to abstract EEG representations from low-level fluctuations to high-level cognitive patterns. Within each block, separable convolutions are used to capture temporal, spatial, and spectral information, while self-attention and cross-attention mechanisms facilitate long-range and inter-branch interactions.

Between stages, temporal resolution is reduced through average pooling, and channel dimensions are adjusted via pointwise convolutions (see the DownSampling module), facilitating effective hierarchical representation learning. This pyramidal architecture, inspired by designs such as the Swin Transformer, supports efficient multi-scale modeling while retaining both fine-grained local cues and global contextual information. The outputs from all stages are concatenated along the token dimension and passed through a final multi-head attention layer, which refines global interactions before feeding into the classification head.

To further enhance training stability and improve temporal sensitivity, HCFT integrates learnable positional embeddings and optionally incorporates a lightweight normalization strategy known as Dynamic Tanh Normalization. DyT serves as a task-adaptive alternative to conventional LayerNorm, introducing a learnable nonlinear transformation that can stabilize information propagation across deep layers without increasing inference overhead. In practice, either DyT or LayerNorm can be selected depending on the characteristics of the dataset, allowing the framework to balance generalization and stability across diverse decoding scenarios.

\subsection{CFT Block}
The core innovation of the HCFT architecture lies in its modular unit, termed the CFT Block, which comprises two integrated components: a dual-branch convolutional feature extractor and an attention-based fusion module. This design enables the model to capture complementary fine-grained local patterns and long-range global dependencies from EEG signals across multiple semantic levels.

\subsubsection{Dual-Branch Convolutional Feature Extractor}
 To extract EEG features from distinct perspectives, we construct a parallel dual-branch structure. Given an input EEG tensor  $\textbf{X} \in\mathbb{R}^{H\times W\times C}$, where \textit{B} denotes the batch size, \textit{C} the number of channels, and \textit{T} the temporal length, two feature extraction pathways are applied in parallel:

\textbf{Temporal Convolution Branch (1D)}: This branch treats the EEG input as a multi-channel time series data and employs depthwise separable 1D convolutions to capture temporal dynamics at the channel level. Specifically, a depthwise convolution filters each channel independently, followed by a pointwise convolution to enable inter-channel fusion and dimensional projection. Batch normalization, GELU activation, and dropout are applied for regularization and stability. The computation can be expressed as:
`
\begin{equation}
\textbf{{T}}_1 = \operatorname{GELU}(\operatorname{BN}(\operatorname{DWConv1D}(\operatorname{\textbf{X}})))
\label{eq}
\end{equation}

\begin{equation}
\textbf{{Y}}_1 = \textbf{{T}}_1 = \operatorname{GELU}(\operatorname{BN}(\operatorname{PWConv1D}(\operatorname{\textbf{T}_1})))
\label{eq}
\end{equation}

where \textit{P} is the number of patches and \textit{D} is the embedding dimension.

\textbf{Spatiotemporal Convolution Branch (2D)}: This branch considers EEG as a 2D array $\textbf{X}' \in\mathbb{R}^{B\times 1 \times C\times T}$ in treating channels and time jointly. First, a temporal convolution with kernel shape scans along the time axis across all channels:

\begin{equation}
\textbf{{TS}}_1 = \operatorname{GELU}(\operatorname{BN}(\operatorname{Conv2D}(\operatorname{\textbf{X}})))\in \mathbb{R}^{B\times F \times C \times T'}
\label{eq}
\end{equation}

Next, a depthwise spatial convolution (kernel size  ) is applied to compress the channel dimension:

\begin{equation}
\textbf{{TS}}_2 = \operatorname{GELU}(\operatorname{BN}(\operatorname{DWConv2D}(\operatorname{\textbf{X}})))\in \mathbb{R}^{B\times F \times C \times T'}
\label{eq}
\end{equation}

Then, a pointwise convolution reshapes the output to match the temporal branch format.

\begin{equation}
\textbf{{TS}}_2 = \operatorname{GELU}(\operatorname{BN}(\operatorname{DWConv2D}(\operatorname{\textbf{X}})))\in \mathbb{R}^{B\times F \times C \times T'}
\label{eq}
\end{equation}

Finally, the 4D tensor is rearranged and flattened to match the temporal branch output format:$\textbf{Y}_2 \in\mathbb{R}^{B\times P \times D}$, with D consistent with the other branch and format:$\textbf{T}' = \textbf{P}$.

\subsubsection{Attention-based Fusion Module}
This module receives the outputs $\textbf{Y}_2 \in\mathbb{R}^{B\times P \times D}$ from both branches and processes them through three steps: \textbf{Self-Attention Enhancement}, \textbf{Cross-Attention Fusion}, and \textbf{Feedforward Network (FFN)}. The design is inspired by conditional attention mechanisms used in DiT, where low-dimensional signals guide the modeling of main modality representations. Following this principle, we treat the output of the temporal branch as the conditioning signal and the output of the spatiotemporal branch as the main representation, enabling temporal cues to guide spatial-contextual feature refinement. We assume the outputs are equal-length sequences: $\textbf{Z}_{cond}, \textbf{Z}_{main} \in\mathbb{R}^{B\times P \times D}$.

\textbf{Self-Attention Enhancement}: To enable long-range temporal dependencies within the spatiotemporal branch, a standard MHSA mechanism is applied. A Pre-Norm structure is adopted, where the normalization layer can be configured as either Dynamic Tanh Normalization (DyT) or conventional LayerNorm, depending on the task setting. In this work, we explore both variants across datasets, allowing the model to flexibly balance stability and generalization.

\begin{equation} \tilde{\textbf{Z}}_{\text{main}} = \operatorname{Norm}(\textbf{Z}_{\text{main}}), \quad \operatorname{Norm} = \operatorname{DyT}/\operatorname{LN} \label{eq:6} \end{equation}
Then we project to query, key, and value:
\begin{equation}
\textbf{Q}_s = \textbf{W}_q^{(s)} \tilde{\textbf{Z}}_{\text{main}},\;
\textbf{K}_s = \textbf{W}_k^{(s)} \tilde{\textbf{Z}}_{\text{main}},\;
\textbf{V}_s = \textbf{W}_v^{(s)} \tilde{\textbf{Z}}_{\text{main}}
\label{eq:7}
\end{equation}

The self-attention result is:

\begin{equation} 
\textbf{A}_{\text{self}} = \operatorname{MHSA}(\textbf{Q}_s, \textbf{K}_s, \textbf{V}_s) \in \mathbb{R}^{B \times P \times D},\;
\quad \textbf{Z}_1 = \textbf{A}_{\text{self}} + \textbf{Z}_{\text{main}} \label{eq:8} 
\end{equation}

\textbf{Cross-Attention Fusion}: The temporal branch captures local dynamics while the spatiotemporal branch encodes structural and contextual information. After normalization:
\begin{equation} \tilde{\textbf{Z}}_{\text{1}} =\operatorname{DyT}(\textbf{Z}_{\text{1}}) \end{equation}

We project the temporal branch as query and the main branch as key and value:
\begin{equation}
\textbf{Q}_\text{c} = \textbf{W}_q^{(c)} \textbf{Z}_{\text{cond}},\;
\textbf{K}_\text{c} = \textbf{W}_\text{k}^{(c)} \textbf{Z}_{\text{cond}},\;
\textbf{V}_\text{c} = \textbf{W}_\text{v}^{(c)} \textbf{Z}_{\text{cond}},\;
\label{eq:7}
\end{equation}
Then compute with the cross-attention (CSA) operation:
\begin{equation} 
\textbf{A}_{\text{cross}} = \operatorname{CSA}(\textbf{Q}_c, \textbf{K}_c, \textbf{V}_c),\;
\quad \textbf{Z}_2 = \textbf{A}_{\text{cross}} + \textbf{Z}_{\text{1}}
\end{equation}
Feedforward and Final Integration: An FFN is applied, and residual and conditional inputs are integrated:
\begin{equation} 
\textbf{Z} = \operatorname{FFN}(\text{DyT}(\textbf{Z}_2)+\textbf{Z}_2+\textbf{Z}_\textbf{cond}
\end{equation}
\textbf{Position Encoding and DyT}: A learnable positional encoding is added in the main branch to encode sequence order. DyT (Dynamic Tanh) replaces LayerNorm to improve training stability. It scales the input with a learnable factor $\alpha$:
\begin{equation}
\quad \operatorname{DyT}(x) = \operatorname{tanh}(\alpha  \cdot x ) +\textbf{Z}_1
\end{equation}
\subsection{Multi-Scale Aggregation and Final Classification}
Within each stage, multiple CFT Blocks are stacked to expand the model’s effective receptive field while maintaining consistent patch length and embedding dimension. Between stages, average pooling is applied along the temporal axis to progressively reduce resolution, and pointwise convolution is used to align the channel dimensions across scales:
\begin{equation} 
\textbf{S}_{i+1} = \operatorname{AvgPooling}((\textbf{S}_i),i=1,2,3,4
\end{equation}
Each stage output  is projected to the final dimension space and concatenated:
\begin{equation} 
\textbf{Z}_\text{final} = \operatorname{Concat}(\textbf{S}_1',\textbf{S}_2',\dots,\textbf{S}_n')
\end{equation}
A final MHSA module is applied to globally refine the fused representation, followed by normalization (LayerNorm or DyT), global average pooling, and a fully connected layer for classification (trained using cross-entropy loss):
\begin{equation} 
y = \operatorname{FC}(\text{GAP}(\text{Norm}(\text{MHA}(\textbf{Z}_\text{final})))
\end{equation}
In summary, the proposed hierarchical convolutional fusion Transformer HCFT is a specialized deep learning framework for EEG decoding. It harnesses a dual-branch convolutional encoder to capture complementary fine-grained temporal and spatiotemporal features. These features are dynamically integrated by the cross-attention and hierarchically aggregated within a multi-stage encoder, which employs a shared patch structure and a multi-scale fusion strategy. Consequently, the architecture successfully models global long-range dependencies while simultaneously preserving critical local details.

\section{DATASET AND EXPERIMENTAL SETUP}
To comprehensively evaluate the modeling capacity and architectural effectiveness of the proposed HCFT framework, we conduct empirical studies on two widely used and representative EEG datasets: BCI Competition IV-2b for MI classification and CHB-MIT for seizure prediction. These datasets differ significantly in task setting, subject population, and signal characteristics, thereby providing a rigorous testbed for assessing the generalizability and robustness of our method.

\subsection{Datasets and Preprocessing}
\textit{\textbf{Dataset I—BCI Competition IV 2b}}:  This dataset comprises 2-second EEG trials of left- and right-hand motor imagery from nine healthy subjects, recorded at 250 Hz using three electrodes (C3, Cz, and C4). To evaluate cross-subject generalization, we adopt a leave-one-subject-out (LOSO) protocol: in each of the nine folds, data from eight subjects are used for training, and the held-out subject is used for testing. All trials are z-score normalized on a per-channel basis, and fixed-length 2-second segments are used as model input. Final performance is reported as the average accuracy and Cohen’s kappa across the nine folds.

\textit{\textbf{Dataset II—CHB-MIT Scalp EEG}}:  The CHB-MIT dataset comprises continuous, multi-channel intracranial EEG recordings from pediatric subjects with epilepsy, sampled at 256 Hz. We adhere to the preprocessing pipeline established in \cite{16}, which involves channel selection, bandpass filtering, and data segmentation for a seizure prediction task. From the original 23 channels, we retain 18 standardized bipolar channels. The study includes data from 20 subjects, having excluded subjects 04, 12, 13, and 24 due to excessive recording durations, inconsistent channel configurations, or inadequate interictal data. The continuous EEG is segmented into 5-second epochs. The preictal class is defined as epochs occurring within the 30-minute window preceding a seizure onset. Interictal samples are drawn from seizure-free periods, with explicit exclusion of ictal activity, a 4-hour buffer zone around each seizure, and a 30-second postictal window to prevent transitional contamination. All data segments are subsequently z-score normalized per channel.

Noise artifacts are suppressed using a sixth-order Butterworth band-stop filter targeting powerline harmonics (57–63 Hz and 117–123 Hz), followed by a 1 Hz high-pass filter to eliminate baseline drift. The original 256 Hz sampling rate is preserved throughout preprocessing. A subject-wise 70/30 split is used for training and testing, respectively.

\subsection{Experiment Setting}
\subsubsection{Performance Metrics}
To ensure a comprehensive evaluation across diverse EEG decoding scenarios, we employ a suite of established metrics tailored to each task’s specific requirements. For general classification tasks, we report Accuracy, the standard deviation (Std) of accuracy across subjects, and Cohen’s Kappa coefficient (Kappa). Accuracy quantifies the proportion of correctly classified samples, and Std measures cross-subject performance consistency. Kappa further assesses inter-rater reliability by accounting for chance agreement, making it robust to class imbalance.

In seizure detection tasks using the CHB-MIT dataset, we adopt domain-specific metrics: Sensitivity (Sens), Specificity (Spec), the False Positive Rate per hour (FPR/h), and the Area Under the ROC Curve (AUC). Sensitivity and Specificity evaluate the model’s precision in identifying seizure and non-seizure events, respectively. The FPR/h is a critical clinical indicator quantifying hourly false alarms, while the AUC provides a holistic view of the model’s discriminative capability across all classification thresholds.

\subsubsection{Implementation Details}
 To ensure robust training and fair cross-dataset comparisons, the proposed HCFT model was implemented in Python 3.11 using PyTorch 2.0. All experiments were conducted on a single NVIDIA GeForce RTX 4090 GPU. The model was trained using the AdamW optimizer with an initial learning rate of 0.001 and a weight decay of 0.00125. A cosine annealing scheduler with a maximum cycle length of 32 epochs was used for learning rate adjustment. The optimizer’s momentum coefficients were set to (0.675, 0.999). Training was conducted for up to 250 epochs with early stopping based on validation performance. A batch size of 64 was used for all experiments.

\subsection{Model Parameters}
TABLE I lists the detailed model parameters. Specifically, the internal structure of HCFT is designed to match the temporal and spectral characteristics of EEG. Both the 1D temporal branch and the 2D spatiotemporal branch use convolutional kernels of length 15, effectively covering the 8–30 Hz $\alpha / \beta$  rhythm band that is critical for cognitive decoding. Depthwise separable convolutions are applied independently across EEG channels to extract localized rhythms, followed by 1×1 pointwise convolutions that perform cross-channel fusion and projection into a unified embedding space with dimension D.

\begin{table}[H]
    \centering
    \begin{threeparttable}
    \caption{Module and Parameters of the Model}
    \begin{tabular}{ccccc}\toprule
         Module&  Input Shape&  Output Shape&  Kernel& Stride\\\midrule
         2D Conv&  (C, T)&  (N, C, T)&  (1,15)& (1,1)\\
         2D DW Conv&  (N, C, T)&  (N, C, T’)&  (C,1)& (1,1)\\
         2D Pw Conv&  (N, C, T’)&  (P, D)&  (1,1)& (1,1)\\
         1D Dw Conv&  (C, T)&  (N, T)&  15& 1\\
         1D Pw Conv&  (N, T)&  (P, D)&  1& 1\\
         Pooling (S1$\sim$S3)&  (P, D)&  (P', D)&  (1, 10)& (1,2)\\
         Pooling  (S4)&  (P, D)&  (P'', D)&  (1,4)& (1,2)\\
         Pooling  (S-Concat)&  (P, D)&  ($P_{\text{final}}$, D)&  (1,75)& (1,15)\\ \bottomrule
    \end{tabular}
    \begin{tablenotes}
        \footnotesize
        \item *C: EEG channel count, T: original time steps, N indicates intermediate feature channels, P represents patch numbers (token number), and D is the embedding dimension.
    \end{tablenotes}
    \end{threeparttable}
    \label{table I}
\end{table}

Between hierarchical stages, we introduce staged average pooling to progressively adjust the temporal resolution. Stages 1–3 (i.e., S1~S3 in TABLE I apply average pooling with a kernel of (1, 10) and a stride of (1, 2), halving the sequence length at each stage to expand the receptive field while preserving essential temporal details. Stage 4 (i.e., S4) further compresses the representation using a kernel of (1, 4) and stride of (1, 2). After concatenating the outputs of all stages, a long context pooling operation with a kernel of (1, 75) and stride of (1, 15) uniformly compresses the multi scale sequence into a fixed length, thereby satisfying the input requirements of subsequent attention modules and the classification head.

Finally, the embedding dimension D and the number of attention heads are chosen to balance representational capacity and computational efficiency. We set D = 32 and employ two attention heads, which in small token scenarios allows for parallel modeling of diverse patterns while keeping computational overhead within acceptable bounds. Through these carefully calibrated parameter choices, HCFT remains efficient, yet fully captures the multi scale spatiotemporal dependencies inherent in EEG data.

\section{RESULTS}
This section presents a comprehensive evaluation of the proposed HCFT. We first benchmark its performance against state-of-the-art (SOTA) methods on two public EEG datasets. Subsequently, we perform ablation studies to isolate and quantify the contribution of each core component. Finally, we provide visual analyses to offer interpretable insights that substantiate the model’s design rationale.
\subsection{Head-to-head Comparison Results}
TABLE \ref{tab:table 2} summarizes the cross subject performance of the proposed model and 15 representative baselines on Dataset I. Overall, the proposed HCFT attains a mean accuracy of 80.83\%, which is the highest among all contenders with small standard deviations. The performance advantage of HCFT is consistent and statistically significant. It outperforms canonical CNN baselines with ConvNet by 10.18\% (p<0.01), EEGNet by 7.38\% (p<0.05), and MSNN by 5.81\% (p<0.05), respectively. A further comparison with the CNN Transformer hybrids(Hybrid-s/t CViT) shows that the gains increase to 16.39\% and 14.04\% (both p < 0.01). Furthermore, HCFT always maintains a 4~6\% accuracy advantage over recent Transformer-based models like CTNet, ConTraNet, EEGPT, and MSCFormer. More specifically for the individual, while HCFT was marginally outperformed on subjects S3, S5, and S6 by EEGNet, CTNet, and MSNN respectively, it secured the top accuracy on six out of nine subjects (S1, S2, S4, S7, S8, S9). Overall, HCFT demonstrated a clear and comprehensive performance lead, surpassing all 15 contenders and outperforming five models (MSHCNN, Conformer, EEGCCT, Hybrid EEGNet, MSVTNet) on every single subject.

In terms of the robustness, HCFT demonstrates superior stability with a cross-subject standard deviation of 7.61\%. This is markedly lower than models like SCNN (over 12\%) and represents a 2-3\% improvement over other contenders, indicating that HCFT is less sensitive to inter-individual variability and delivers more consistent performance. Further, regarding classification agreement, HCFT achieves a Cohen’s $\kappa$ of 0.6165, signifying substantial agreement and surpassing

\begin{table*}[h]
    \captionsetup{font=footnotesize,justification=centering}
    \caption{\\Classification Accuracy And Kappa Values of Different Methods on Dataset I [Avg Acc: The Average Accuracy (\%); S1~S9 Means Subject 1~9 Involved in The Dataset I; ‘-’ Indicates that the exact values were not reported by the original paper; *: P$<$0.05, **: P$<$0.01.]}\label{tab:table 2}
    \centering
    \footnotesize
    \resizebox{\textwidth}{!}{
    \begin{tabular}{cccccccccccccc}
        \toprule
        \textbf{Year} & \textbf{Methods} & \textbf{S1} & \textbf{S2} & \textbf{S3} & \textbf{S4} & \textbf{S5} & \textbf{S6} & \textbf{S7} & \textbf{S8} & \textbf{S9} & \textbf{Avg Acc} & \textbf{Std} & \textbf{Kappa} \\
        \midrule
        2017 & ConvNet\cite{17}    & 64.19 & 62.9 & 67.58 & 72.06 & 75.87 & 72.01 & 81.51 & 79.02 & 60.68 & 70.65** & 7.33 & 0.4134 \\
        2019 & EEGNet \cite{18}     & 66.15 & 71.08 & \textbf{72.01} & 56.48 & 80.24 & 78.78 & 85.03 & 79.54 & 71.74& 73.45** & 8.64 & 0.4684 \\
        2019 & MSNN \cite{19}  & 74.72	& 65.29	 & 57.63	 & 91.21 & 	74.72	 & \textbf{85.55}	 & 72.91	 & 76.57	 & 76.66	 & 75.02*	 & 9.88	 & -  \\
        2020 & Hybrid s-CViT \cite{20}   &   68.47	 & 56.91	 & 50.42 & 	81.08	 & 60.68	 & 61.67 & 	62.22	 & 70.00	 & 68.47 & 	64.44**	 & 8.81 & 	- \\
        2021 & Hybrid t-CViT \cite{20}       &  66.39	 & 55.74 & 	52.36	 & 82.7	 & 72.57	 & 63.89	 & 68.89	 & 65.92	 & 72.64	 & 66.79**	 & 9.12	 & - \\
        2022 & MSHCNN \cite{21}     & 76.80&	66.32	&57.36	&91.75	&79.59&	82.63	&74.16	&80.13&	75.55&76.03**	&9.79&	- \\
        2023 & Conformer \cite{8}     &   65.89	 & 64.43	 & 67.45	 & 84.45	 & 72.24	 & 76.56	 & 77.86	 & 69.23	 & 74.87	 & 76.4**	 & 6.51	 & 0.4521 \\
        2024 & EEGCCT \cite{22} &  68.75	& 59.6	& 59.9	& 89.21& 	73.44	& 75.39& 	76.3	& 75.76& 	77.73	& 73.26** & 	9.21	& 0.4587 \\
        2024 & Hybrid EEGNet \cite{23}    &  71.53&	65.00	&58.75	&84.86&	78.78	&77.50	&77.92	&73.68	&75.41	&73.72**&	7.82	&- \\
        2025 & CTNet \cite{24}  &  76.25	  &71.03	  &66.39  &81.76	  &\textbf{83.11}	  &77.22	  &79.17	  &73.56	  &77.92	  &76.27*	  &5.26	  &0.5252 \\
        2025 & EEGPT \cite{25}     & 72.22 &	69.71	 &61.53 &	78.78 &	81.08	 &70.42 &	83.89 &	83.82	 &70.83	 &74.70**	 &7.61	 &0.4936 \\
        2025 & SCNN \cite{27}  & 	- & 	  -	 & -	 & -	 & -	 & -	 & - & -	 & -	 & 79.38	 & 14.17	 & - \\
        2025 & MSCFormer \cite{28}   & 76.11	&71.18	&62.36	&81.35	&81.08	&74.72	&78.89&	76.18	&75.42	&75.25**	&5.80	&0.5051\\
        2025 & ConTraNet \cite{29} &   72.92	&72.94&	63.75	&83.51	&82.70	&80.69	&84.44	&77.37	&70.83	&76.57**	&6.97	&- \\
        2025 & \textbf{HCFT}     & \textbf{78.62}	&\textbf{73.23}	&67.71	&\textbf{93.92}	&82.72	&82.68	&\textbf{86.17}	&\textbf{84.47}	&77.94	&\textbf{80.83}	&7.61	&\textbf{0.6165} \\
        \bottomrule
    \end{tabular}
    }
\end{table*}

the best values reported by CTNet (0.5252). It is important to note that although we reproduced the published code for ConvNet, EEGNet, Conformer, EEGCCT, CTNet, MSVTNet, EEGPT, and MSCFormer to obtain $\kappa$, such the metric was not reported in other compared literatures, thus limiting a more comprehensive benchmark.

We further evaluate HCFT on the Dataset II and compare it with a series of SOTA seizure prediction methods reported from 2018 to date. The comparative results, as summarized in TABLE \ref{table 3}, include well-known baselines such as CNN, TTT, TGCNN, BSDCNN, MAAE, and more recent frameworks including DRSN-GRU, MCNN, ASRU-RW, DWT, STCNN, HviT-DUL, M-d-C, MTLG, B2-ViT, MSAN and NSFA-Net. These methods vary widely in terms of protocol design, such as the number of selected patients, evaluation strategies (e.g., LOSO, subject-specific splits), and employed metrics.

Despite these diversities, HCFT achieves the highest overall performance, attaining a sensitivity of 99.10\%, a false positive rate of only 0.0236/h, and a specificity of 98.82\%, while being evaluated on all available patients under a leave-one-out scheme. This performance surpasses most previous models, including the recent ASRU-RW in 2023 (98.96\%, 0.048/h), MTLG in 2024 (98.24\% sensitivity, 0.033/h FPR), and NSFA-Net in 2025 (98.68\%, 0.038/h), which often excluded specific patient subsets or relied on partial data splits.

Taken together, these results demonstrate that HCFT offers superior performance in the challenging cross subject, long-sequence EEG decoding task, thereby underscoring its strong potential for BCI applications.

\begin{table*}[!htbp]
    \captionsetup{font=footnotesize,justification=centering}
    \caption{Comparison Results of Different Methods on Dataset II for Seizure Prediction (SOP: Seizure Occurrence Period; SPH: Seizure Prediction Horizon; EP: Excluded Patients; LOOCV: Leave-One-Out Cross-Validation; “–” Indicates Values Not Reported by the Paper)}
    \centering
    \footnotesize
    \resizebox{\textwidth}{!}{
    \begin{tabular}{ccccc}\toprule
        \textbf{Year}  & \textbf{Method}&  \textbf{Results}&  \textbf{SOP \& SPH (Minutes)}& \textbf{Notes}\\\midrule
        2018 &  CNN \cite{30} & SENS: 81.20\%, FPR:0.16/h &	30 \& 5	& EP: 04, 06~08, 11, 12, 15~17, 22\\
        2022 &  TTT \cite{16} & SENS: 96.01\%; FPR: 0.047/h	& 30 \& 3	& EP: 12, 13, 15 \\
       2022 &  TGCNN \cite{31} & SENS: 91.5\%; FPR: 0.145/h; AUC: 0.935	& 30 \& 5	& EP: 04, 06, 07, 11, 12, 15, 22 \\
        2022 &  BSDCNN \cite{32} & SENS: 94.69; FPR: 0.095/h; AUC:0.970 & 30 (PIL) \& 5	& With 01, 05, 08, 10, 14, 22 \\
       2022 &  MAAE \cite{33} & SENS: 82.20\%; FPR: 0.13/h; AUC: 0.940	& - & 	LODO (Leave-One-Domain-Out), with 01, 03, 05, 06, 08, 10, 13, 14, 18, 20  \\
        2023 &  DRSN-GRU \cite{33} & SENS: 90.54\%; FPR: 0.11/h; AUC: 0.880	& 30 \& 5	& LOOCV, with 01, 02, 03, 16 \\
       2023 &  MCNN \cite{35} & SENS: 82.00\%; FPR:0.058/h	& - &	All patients \\
        2023 & ASRU-RW \cite{36} & SENS: 98.96; FPR: 0.048/h	& 30 \& 5	& LOOCV (based on the subject-specific, 80\% train, 20\% test), EP: 12, 15 \\
        2023 & DWT \cite{37} & SENS: 91.7\%; \textbf{FPR: 0/h}	& 25 \& 5	& EP: 12, 13, 24 17 for train, 4 for test (04, 10, 06, 19) \\ 
        2023 &  STCNN  \cite{38} & SENS: 92.94\%; FPR: 0.073/h; AUC: 0.9596	& -	& All Patients \\
        2023 & HviT-DUL \cite{39} & SENS: 87.90\%; FPR: 0.056/h; AUC: 0.9370	& 30 \& 1	& EP: 04, 06, 07, 12, 15 \\
        2024 & M-d-C \cite{40} & SENS: 97.80\%; FPR: 0.059/h AUC: 0.9350	& -	& LOOCV \\
        2024 & MTLG \cite{41} & SENS: 98.24\%; FPR: 0.033/h	& 30 \& 5	& EP: 04, 12, 13, 24 \\
        2024 & B2-ViT \cite{41} & SENS: 93.30\%; FPR: 0.057/h; AUC: 0.9230	& 30 \& 5	& EP: 04, 06, 07, 08, 11, 12, 15, 16, 17, 22 \\
        2025 & MSAN \cite{43} & SENS:96.27\%; \textbf{FPR: 0/h}; AUC:0.9300	& 60 \& 5	& EP: 04, 12, 15, 19 \\
        2025 & NSFA-Net \cite{44} & SENS: 98.68\%; FPR: 0.038/h; AUC:0.9060	& 30 \& 5	& EP: 04, 12, 15, 19 \\
        2025 & \textbf{HCFT}	& \textbf{SENS: 99.10\%}; FPR: 0.0236/h;\textbf{SPEC: 98.82\%}	& 30 \& 3	& EP: 04, 12, 13, 24 \\ 
    \bottomrule
    \end{tabular}
    }
    \label{table 3}
\end{table*}

\subsection{Structural and Task Adaptability Analysis}
The effectiveness of architectural choices and parameters in EEG decoding can vary significantly across different tasks. To rigorously evaluate the robustness and generalizability of the proposed HCFT model under such heterogeneous conditions, we conduct extensive parameter sensitivity and module-level ablation studies on two representative datasets.

\begin{figure}[!b]
    \centering
    \includegraphics[width=1\linewidth]{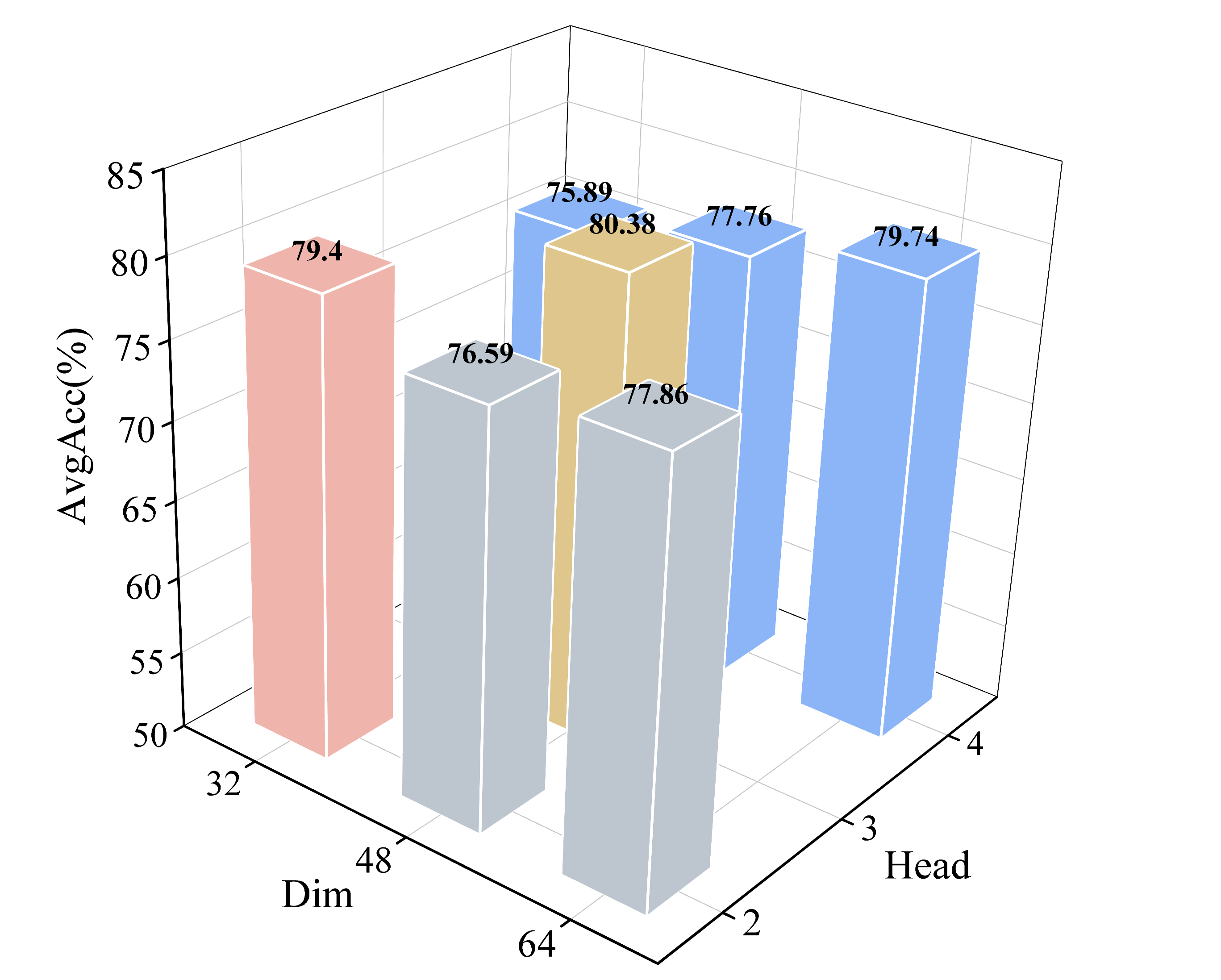}
    \caption{Accuracy effect of embedding dimension and number of heads.}
    \label{fig 2}
\end{figure}

This evaluation is structured in two phases: First, a systematic sensitivity analysis on Dataset I (MI classification) identifies the optimal configuration for key hyperparameters, including encoder depth, embedding dimension, and the number of attention heads. Second, to assess cross-task adaptability, this optimal configuration is directly transferred to Dataset II (epileptic seizure prediction), where targeted ablation experiments quantify the contribution of each core structural component.

\subsubsection{ Parameter Sensitivity Analysis }

To examine the role of dimensionality and multi-head configurations, a sensitivity analysis on the embedding dimension (D) and the number of attention heads (H), increasing D from 32 to 64 while adjusting H to maintain a consistent per-head dimension. Since the dimension-to-heads ratio must be an integer, configurations with dimensions of 32 or 64 are incompatible with 3 heads and thus being excluded. As illustrated in Fig. \ref{fig 2}, the configuration with D = 48 and H = 3 achieved the optimal results, delivering the highest accuracy with reduced performance variance. However, while increasing D provided marginal gains, it incurred a substantial increase in model complexity. Indeed, as concluded in TABLE \ref{table IV}, with increasing in the embedding dimension D and the number heads H, both the model size (\#Param) and FLOPs significantly increase. Therefore, for an optimal balance of performance and computational efficiency, we selected D = 32 with H = 2 as the default configuration.

\begin{table}[!htbp]
    \centering
    \caption{Computational Efficiency for Specific Dim (D) and Head (H) Configuration with Model Stage Layer N of 1-1-2-1 on Dataset I}
    \label{table IV}
    \footnotesize
    \resizebox{\columnwidth}{!}{
    \begin{tabular}{ccccc}
        \toprule
        \textbf{D} & \textbf{H} &\textbf{ Performance Metric} & \textbf{\#Param} &\textbf{FLOPs} \\
        \midrule
        32 & 2 & Avg Acc: 79.40\% & 88.987K & 72.880M \\
        48 & 3 & Avg Acc: 79.40\% & 194.123K & 161.039M \\
        64 & 4 & Avg Acc: 79.74\% & 339.707K & 283.700M \\
        \bottomrule
    \end{tabular}
    }
\end{table}

 \begin{figure}[!b]
     \centering
     \includegraphics[width=1\linewidth]{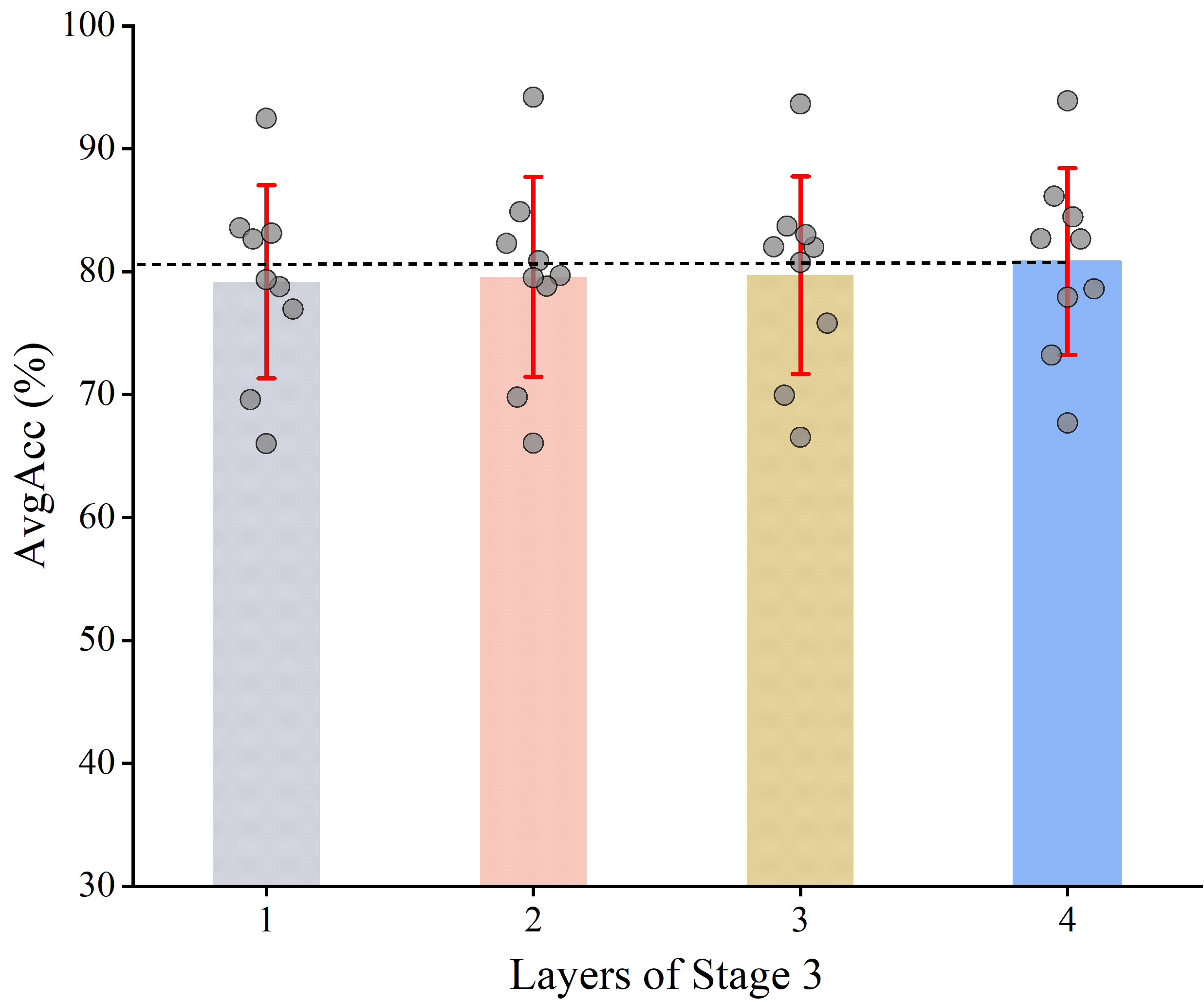}
     \caption{Influence of number of Stage 3 layers on model performance.}
     \label{fig 3}
 \end{figure}

To investigate the impact of model depth, we systematically varied the number of encoder layers while maintaining the above default configuration. We focused depth adjustments on Stage 3 of the architecture, as this stage is responsible for integrating spatiotemporal features from earlier layers and modeling higher-level semantic dependencies. Increasing its depth enhances the model’s capacity to capture complex long-range interactions and inter-channel relationships. In fact, experimental results confirm that deeper configurations yield consistent performance gains. The model stage layer from N1~N4 in Fig. \ref{fig 1} being 1-1-4-1 structure achieved the highest accuracy and Kappa score (see in Fig. \ref{fig 3}), demonstrating that a moderate increase in depth at this critical stage effectively strengthens the model’s ability to learn hierarchical representations without inducing overfitting.

Normalization is critical for ensuring training stability and generalization in Transformer-based models. To evaluate the task-specific efficacy of our proposed Dynamic Tanh Norma-lization of DyT, we compared it against traditional LayerNorm across two distinct EEG decoding paradigms.

The results reveal a clear task-dependent performance trade-off. On the motor imagery classification task (Dataset I), DyT achieved a +2.01\% higher average accuracy than LayerNorm (TABLE \ref{table v}). We attribute this advantage to DyT’s dynamic, learnable scaling followed by a saturating nonlinearity and thus exceling at highlighting transient, event-driven features in non-stationary MI-EEG signals, unlike LayerNorm’s reliance on global statistics.

Conversely, LayerNorm proved more effective for the seizure prediction task (Dataset II). This task involves signals with higher continuity, stronger rhythmicity, and more pronounced long-range dependencies, where LayerNorm’s global statistical stabilization better handles session-wise variability and ensures robust training. While DyT resulted in a slight performance decrease in this context, it consistently reduced model size (\#Param) and computational cost (FLOPs), underscoring its value for lightweight deployment.

Consequently, we adopt DyT for Dataset I and retain LayerNorm for Dataset II. This selective configuration aligns the normalization strategy with the intrinsic statistical properties and temporal dynamics of each specific EEG decoding paradigm.

\begin{table}[h]
    \centering
    \caption{Evaluation of Parameter Complexity}
    \label{table v}
    \footnotesize
    \begin{tabularx}{\columnwidth}{c c >{\centering\arraybackslash}X c c}
        \toprule
        \textbf{Strategy} & \textbf{Dataset} & \textbf{Performance Metric} & \textbf{\#Param} & \textbf{FLOPs} \\
        \midrule
        LN & I & Avg Acc: 78.82\% ± 7.13 & 88.987K & 72.880M \\
        LN & II & SENS: 99.1\%; SPEC: 98.82\%; FPR: 0.0236/h & 194.123K & 161.039M \\
        DyT & I & Avg Acc: 80.83\% ± 7.61 & 339.707K & 283.700M \\
        DyT & II & SENS: 96.34\%; SPEC: 98.42\%; FPR: 0.0268/h & 339.707K & 283.700M \\
        \bottomrule
    \end{tabularx}
\end{table}

\subsubsection{Ablation Experiment}
: To quantitatively assess the contributions of the core components in the proposed HCFT model, we conducted systematic ablation studies on both Dataset I and Dataset II. The experiments evaluate four critical modules: the self-attention mechanism, the conditional cross-attention module, the stage-wise feature concatenation, and the final multi-head attention layer.
As listed in TABLE \ref{table VI}, ablation results on Dataset I confirm that each architectural component is integral to HCFT’s performance, as the removal of any single module induces consistent degradation. The most substantial performance drop occurred when the cross-attention mechanism was disabled (79.10\% ± 7.97), highlighting its critical role in fusing class-discriminative information from the temporal and spatiotemporal branches. Removing self-attention also caused a notable decline (79.60\% ± 7.96), underscoring its necessity for modeling global temporal dynamics, particularly in $\mu$- and $\beta$-rhythms. Similarly, omitting stage-wise concatenation led to reduced accuracy (79.75\% ± 8.39), affirming the importance of multi-scale hierarchical features for robust decoding.

For Dataset II, we similarly performed ablation experiments by sequentially removing the self-attention, cross-attention, stage concatenation, and final MHSA modules. The results exhibited marginal fluctuations in performance, where the sensitivity remained consistently high ($>$ 98\%) with a false positive rate below 0.03/h. This robustness indicates a degree of structural redundancy within HCFT for long-range EEG

\begin{table*}
    \centering
    \caption{Ablation Results and Performance Impact of Key HCFT Components on Dataset I and II}
    \label{table VI}
    \resizebox{\textwidth}{!}{%
    \begin{tabular}{cccccc}
        \toprule
        \textbf{Self-Attn.} & \textbf{Cross-Attn.} & \textbf{Stages Concat} & \textbf{Final MHA} & \textbf{Dataset I Avg Acc (\%)} & \textbf{Dataset II Performance} \\
        \midrule
        $\times$ & \checkmark & \checkmark & \checkmark & 79.60±7.96  & SENS: 98.31\%; FPR:0.025/h; SPEC:98.47\% \\
       \checkmark &  $\times$ & \checkmark & \checkmark & 79.10±7.97 & SENS: 98.84\%; FPR:0.0240/h; SPEC:97.0\% \\
       \checkmark & \checkmark &  $\times$ & \checkmark & 79.75±8.39 & SENS: 98.78\%; FPR:0.0280/h; SPEC:98.30\% \\
        \checkmark & \checkmark &\checkmark & \checkmark & 79.63±7.76 & SENS: 98.73\%; FPR:0.02512/h; SPEC:98.86\% \\
        \checkmark & \checkmark & \checkmark & \checkmark & 80.83±7.61 &  SENS: 99.10\%; FPR:0.0236/h; SPEC:98.82\%\\
        \bottomrule
    \end{tabular}
    }
\end{table*}

 \begin{figure*}[!b]
    \centering
    \includegraphics[width=\textwidth]{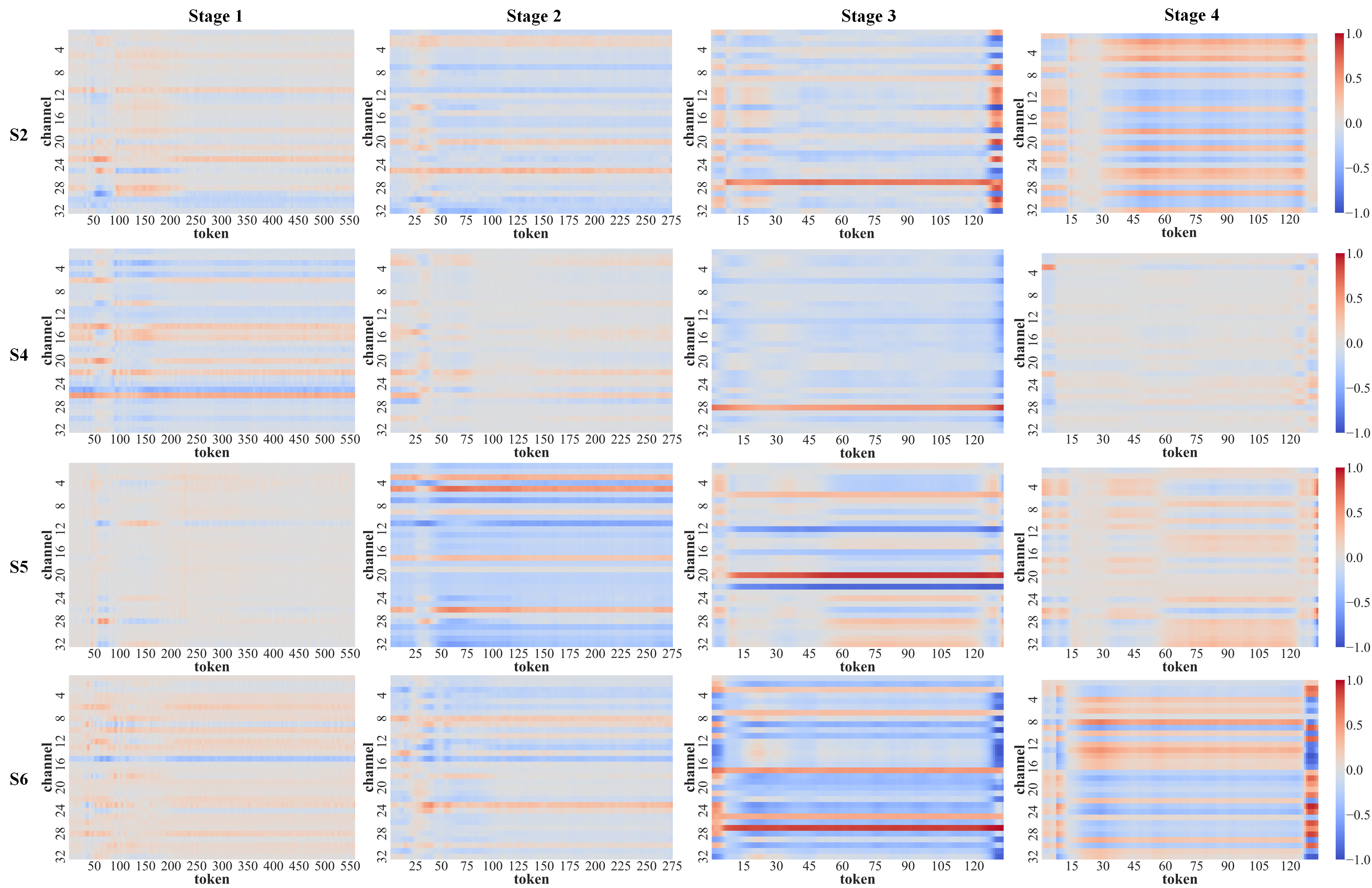}
    \caption{The proposed HCFT framework, include Multiple Stage design with different number of CFT Blocks}
    \label{fig 4}
\end{figure*}

tasks such as seizure prediction. The distinct, pronounced patterns characteristic of preictal activity appear to be captured effectively by the core spatiotemporal modeling, even when specific components are disabled. This property is highly advantageous for clinical deployment, as it suggests the model can be adapted to various computational constraints without compromising critical detection performance. Moreover, the minor contributions of components like DyT or the final MHA in this context may become more critical in noisier environments, pointing to a pathway for future optimization targeted at challenging real-world scenarios.

\subsection{Visualization}

To interpret internal decision-making of the proposed HCFT, we visualized the channel-wise attention distributions across hierarchical stages on four representative subjects. As illustrated in Fig. \ref{fig 4}, each heatmap presents the normalized attention weights assigned by the model to channels and token
at each processing stage (Stage 1~4), offering a dynamic perspective on how spatial representations evolve hierarchically throughout the network.
For these selected subjects (S2, S4, S5, S6), a clear pattern emerges: early stages (e.g., Stage 1) display a diffuse attention pattern, reflecting an initial, broad exploratory phase across all input channels. As processing advances to deeper stages (Stages 3-4), attention sharpens dramatically, converging onto a sparse set of critical channels and tokens. This progression from exploration to specialization suggests the model hierarchically refines its spatial focus to isolate the most discriminative, task-relevant neural signatures for motor imagery task of Dataset I.

This observed sharpening of attention is not merely a result of deeper network compression, but rather reflects the coordinated interaction between stacked convolutional filters and the model’s attention mechanisms. While early layers Combining the heatmap and TABLE \ref{tab:table 2}, four cases (S2, S4, S5, S6) show structured multi-stage attention concentration and are able to learn effective features in the shallow stages and key features in the deeper stages, reflecting advanced decoding performance.
 
provide spatially localized receptive fields, the self- and cross-attention modules progressively integrate information across spatial dimensions and across processing stages. This leads to a reconstruction of more abstract spatial representations that are tuned to class-specific neural dynamics, ultimately yielding stronger decoding performance.

\begin{figure}[!b]
    \centering
    \includegraphics[width=1\linewidth]{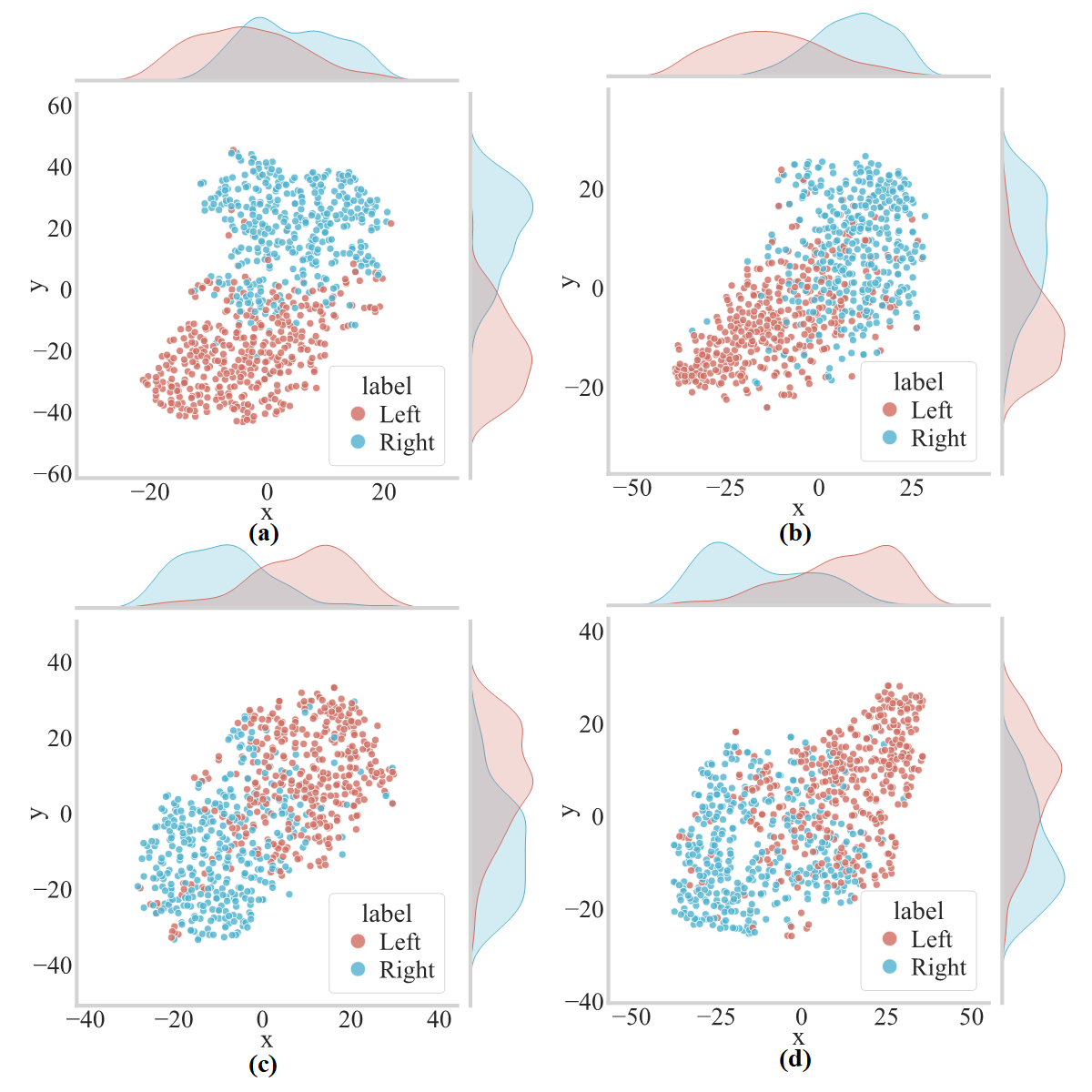}
    \caption{The features for subjects ((a)-S4, (b)-S5, (c)-S7, (d)-S8) of Dataset I by t-SNE visualization.}
    \label{fig 5}
\end{figure}

To further investigate the model’s capability to capture inter-subject variability, we visualized the features for each individual subject. Among the representative subjects (e.g., S4, S5, S7, S8), the t-SNE visualization revealed two well-separated clusters (see in Fig. \ref{fig 5}). It demonstrates that the extracted features exhibit notable intra-class compactness and inter-class separability, validating the model’s ability to discriminate between different categories in the latent space. When combined with the channel-wise token activation maps, a complementary trend emerges: for the subjects S4 and S5 with high performance, attention intensifies progressively across stages, culminating in distinct spatial localization during Stage 3 and 4. This hierarchical refinement implies that the HCFT leverages successive attention layers to suppress irrelevant noise and reinforce salient spatial representations.

In summary, these above presented analyses, including ablation studies, parameter sensitivity evaluations, attention- and feature-based visualizations, collectively confirm the effectiveness and interpretability of the HCFT architecture.

\section{DISCUSSION AND CONCLUSION}
In this study, we proposed the Hierarchical Convolutional Fusion Transformer (HCFT), a novel decoding architecture that addresses the inherent non-stationarity and subject variability of EEG signals through a unified spatiotemporal modeling framework. Inspired by the classical deep learning pipeline of hierarchical feature extraction and multi-scale integration, HCFT incorporates dual-branch convolutions to encode diverse frequency-temporal patterns and employs a hierarchical multi-stage Transformer structure to progressively fuse representations across temporal and spatial dimensions. The attention-based fusion mechanism enhances inter-branch communication, while the inclusion of normalization variants with Dynamic Tanh (DyT) further supports flexibility across tasks with differing signal properties.

The effectiveness of HCFT has been verified on two fundamentally distinct EEG tasks, supporting its applicability across both event-related (i.e., Dataset I) and continuous monitoring (i.e., Dataset II) scenarios. From a structural perspective, the combination of local convolutional priors and global attention-based modeling enables the network to capture both short-term discriminative patterns and long-range dependencies, a design that aligns well with the intrinsic hierarchical organization of neural signals. Moreover, the use of modular stages with shared token structures facilitates deeper feature abstraction while maintaining computational tractability. Attention visualizations further reveal that the model learns to prioritize semantically meaningful channels and time points, offering a degree of interpretability that is often lacking in current deep EEG models \cite{5,13}. 

In summary, HCFT demonstrates highly competitive performance and robust cross-subject generalization, a capability we attribute to its integrated spatiotemporal feature extraction and synergistic use of self-attention and cross-attention mechanisms. This design fosters rich interactions between temporal and spatial representations while effectively balancing local feature extraction with global dependency modeling. Despite this overall strength, HCFT was marginally outperformed on subjects S3, S5, and S6 in Dataset I by specialized architectures. EEGNet’s efficiency in capturing canonical EEG features \cite{18}, CTNet’s enhanced local detail extraction \cite{24}, and MSNN’s multi-scale representational capacity \cite{19} each conferred a subject-specific advantage. These cases highlight that while HCFT provides a powerful generalized framework, the optimal architecture may still vary with individual neurophysiological characteristics.

Meanwhile, the proposed HCFT framework faces some other open challenges that warrant further investigation. A primary limitation is its reliance on dataset-specific hyperparameter tuning for components like the normalization strategy and stage depth, which could impede seamless generalization to novel datasets or subject cohorts. Future work could explore meta-learning or adaptive reparameterization techniques to enable subject-agnostic configuration. Furthermore, the model’s development is constrained by the limited scale and diversity of existing public EEG datasets, which hinders the pursuit of true cross-population robustness. Addressing this fundamental issue will likely require the creation of large-scale, multi-center EEG corpora, potentially facilitated by federated learning frameworks or cross-institutional collaboration frameworks. Finally, while HCFT is designed to be efficient, the computational demands of its multi-stage attention mechanisms remain non-trivial for real-time edge deployment. Future research should prioritize strategies such as hybrid quantization, knowledge distillation, or neural architecture search to achieve a more optimal balance between performance and practical efficiency \cite{5}.

Importantly, HCFT is designed with high extensibility in mind. Despite its compact configuration in the current implementation, the architecture supports straightforward scaling in both depth and width, making it well-suited for future integration with large-scale pretraining paradigms. This provides a promising foundation for developing EEG-based foundation models, where massive cross-task or cross-subject data can be leveraged to build generalizable neural decoders through contrastive learning, masked modeling, or instruction-based pretraining. As the demand grows for flexible and general-purpose neuro intelligent systems, the capacity of HCFT to scale up offers new avenues for robust representation learning and semantic-level EEG interpretation.

Looking ahead, the integration of HCFT with emerging foundation models offers exciting opportunities for the next generation of brain decoding frameworks. Recent developments in large-scale pretrained language and vision models have demonstrated remarkable capacity for generalization under limited supervision. Applying similar paradigms to EEG, such as prompt-based tuning, semantic alignment across modalities, or context-aware reasoning, may enable models that go beyond classification to interpret intention, emotion, or cognition in a more naturalistic manner \cite{45}. Moreover, expanding the decoding framework to incorporate multimodal biosignals such as EMG, eye tracking, or speech could unlock richer representations of human intent and state, bridging the gap between neuroscience and embodied artificial intelligence. These directions will further promote the transformation of EEG decoding from a task-specific pipeline to a flexible, interpretable, and general-purpose neuro-symbolic modeling paradigm.

In conclusion, HCFT offers a principled and extensible approach to EEG decoding by combining structural hierarchy, dynamic fusion, and attention-based abstraction. It provides not only empirical effectiveness across diverse tasks but also a conceptual framework that bridges handcrafted priors and deep representations. As research moves toward more scalable, adaptive, and multimodal decoding systems, HCFT lays a solid foundation for both theoretical exploration and practical advancement in neural signal processing.

\end{document}